\documentclass[twocolumn]{aastex63}

\usepackage{xspace}

\newcommand{\PKS}{PKS 1502+106\xspace}
\newcommand{\TXS}{TXS 0506+056\xspace}

\received{\today}
\submitjournal{ApJ}

\usepackage{booktabs}
\usepackage{multirow}
\usepackage{makecell}

\newcommand{\cf}{{\it cf.}}

\newcommand{\eqs}{Eqs.}

\newcommand{\Fig}{Fig.}

\newcommand{\Sec}{Section}

\newcommand{\Tab}{Tab.}

\newcommand{\tabl}[1]{\Tab~\ref{tab:#1}}

\usepackage[inline]{enumitem}

\graphicspath{{./}{figures/}}

\shorttitle{Multi-wavelength and neutrino emission from  \PKS}
\shortauthors{Rodrigues et al.}

\begin{document}

\title{Multi-wavelength and neutrino emission from blazar \PKS}

\correspondingauthor{Xavier Rodrigues}
\email{xavier.rodrigues@desy.de}

\author[0000-0001-9001-3937]{Xavier Rodrigues}
\affiliation{Deutsches Elektronen-Synchrotron DESY, 15738 Zeuthen, Germany}

\author[0000-0003-2403-4582]{Simone Garrappa}
\affiliation{Deutsches Elektronen-Synchrotron DESY, 15738 Zeuthen, Germany}

\author[0000-0002-5309-2194]{Shan Gao}
\affiliation{Deutsches Elektronen-Synchrotron DESY, 15738 Zeuthen, Germany}

\author[0000-0001-7774-5308]{Vaidehi S. Paliya}
\affiliation{Deutsches Elektronen-Synchrotron DESY, 15738 Zeuthen, Germany}

\author[0000-0002-5605-2219]{Anna Franckowiak}
\affiliation{Deutsches Elektronen-Synchrotron DESY, 15738 Zeuthen, Germany}

\author[0000-0001-7062-0289]{Walter Winter}
\affiliation{Deutsches Elektronen-Synchrotron DESY, 15738 Zeuthen, Germany}



\begin{abstract}

In July of 2019, the IceCube experiment detected a high-energy neutrino from the direction of the powerful blazar \PKS. We perform multi-wavelength and multi-messenger modeling of this source, using a fully self-consistent one-zone model that includes the contribution of external radiation fields typical of flat-spectrum radio quasars (FSRQs). We identify three different activity states of the blazar: one quiescent state and two flaring states with hard and soft gamma-ray spectra. We find two hadronic models that can describe the multi-wavelength emission during all three states: a leptohadronic model with a contribution from photo-hadronic processes to X-rays and gamma rays, and a proton synchrotron model, where the emission from keV to 10~GeV comes from proton synchrotron radiation. Both models predict a substantial neutrino flux that is correlated with the gamma-ray and soft X-ray fluxes. Our results are compatible with the detection of a neutrino during the quiescent state, based on event rate statistics. We conclude that the soft X-ray spectra observed during bright flares strongly suggest a hadronic contribution, which can be interpreted as additional evidence for cosmic ray acceleration in the source independently of neutrino observations. We find that more arguments can be made in favor of the leptohadronic model vis-a-vis the proton synchrotron scenario, such as a lower energetic demand during the quiescent state. However, the same leptohadronic model would be disfavored for flaring states of \PKS if no IceCube events were found from the direction of the source before 2010, which would require an archival search.

\end{abstract}
 


\section{Introduction} \label{sec:intro}

High-energy neutrinos are a unique probe of the high-energy universe capable to identify the acceleration regions of cosmic rays \citep[for recent reviews, see][]{GalloRosso:2018omb,Vitagliano:2019yzm}. A first milestone was the detection of a diffuse neutrino flux with the IceCube neutrino observatory in 2013 \citep{Aartsen:2013jdh}. However, the origin of those neutrinos is still unknown. Active Galactic Nuclei (AGN) are considered promising candidate sources \citep[e.g.][]{1991PhRvL..66.2697S,mannheim92,1993A&A...269...67M,szabo94,1995APh.....3..295M,mastichiadis96,protheroe99,atoyan01,dimitrakoudis12,Murase:2015ndr,Becker:2007sv,Tjus:2014dna}. In particular their relativistic jets are promising sites of cosmic-ray acceleration, which could produce neutrinos in interactions with ambient photon fields or matter in or close to the source. 

To identify possible sources, IceCube has set up a target of opportunity program, which allows the rapid search for multi-wavelength counterparts to high-energy neutrino track events \citep{2017APh....92...30A}. Thanks to this program, the gamma-ray blazar \TXS could be identified as a first compelling neutrino source at the $3\sigma$ level: the high-energy neutrino event IceCube-170922A was found in spatial coincidence with the blazar position and in temporal coincidence with a significant flare in gamma rays \citep{TXS_MM}. An additional excess of lower-energy neutrinos arriving within a 160-day time window in 2014/15 was identified at the $3.5\sigma$ confidence level in an archival search for a time-dependent neutrinos signal from the direction of \TXS \citep{TXS_orphanflare}. However, this neutrino excess was not accompanied by a gamma-ray flare \citep{2019ApJ...880..103G}.

Other possible neutrino blazar associations at lower significance have been pointed out by e.g. \citet{Franckowiak:2020qrq,Giommi:2020hbx,2019ApJ...880..103G, 2018A&A...620A.174K,2016NatPh..12..807K}. Of particular interest is the spatial coincidence of IceCube-190730A with \PKS, which is the 15th brightest gamma-ray source at $>100$~MeV in terms of energy flux among 2863 sources in the fourth catalog of AGN detected by \textit{Fermi}-LAT \citep[4LAC,][]{Fermi-LAT:2019pir}. Given the large redshift of 1.84 \citep{2010MNRAS.405.2302H} the source must have an extremely high intrinsic luminosity. It is also highly variable in the gamma-ray band \citep[see e.g.][]{2010ApJ...710..810A}. While the source did not show an excess in gamma rays during the arrival of IceCube-190730A, the radio flux shows a long-term outburst starting in 2014 and reaching the highest flux density ever reported from this source during the arrival of IC-190730A \citep{2019ATel12996....1K, Franckowiak:2020qrq}, which may indicate a long-term activity of the central engine. IceCube-190730 has an estimated neutrino energy of 300 TeV and a 67\% \textit{signalness\footnote{\url{https://gcn.gsfc.nasa.gov/notices_amon_g_b/132910_57145925.amon}}} based on the procedure by \citet{Blaufuss:2019fgv}.

\PKS~is a broad emission line quasar first identified as a strong radio source in the 178~MHz pencil beam survey \citep{1966MNRAS.132..405C}. At radio frequencies, the source is highly variable and observations from very long baseline interferometry (VLBI) have revealed a core-dominated, one-sided, curved radio jet \citep[e.g.,][]{2004A&A...421..839A}. \PKS~has exhibited large amplitude optical variability \citep[$>$2.5 mag;][]{2008ATel.1661....1M} and the detection of a high degree of optical polarization (up to $\sim$20\%) suggests the dominance of synchrotron emission at these wavelengths. \citet{2011ApJS..194...45S} studied the optical spectrum of this object taken with the Sloan Digital Sky Survey (SDSS) and reported a central black hole mass of 10$^{9.64\pm0.44}M_{\odot}$. In the high-energy gamma-ray band, \PKS~was not detected with Energetic Gamma-Ray Experiment Telescope (EGRET) \citep{1999ApJS..123...79H}. However, a significant $>$100 MeV radiation was detected within the first few months of {\it Fermi}-LAT operation when the blazar was undergoing a huge $\gamma$-ray outburst \citep{2010ApJ...710..810A}. Since then, \PKS~has been extensively studied across the electromagnetic spectrum \citep[cf.][]{2011A&A...526A.125P,2016A&A...586A..60K,2017ApJ...851...33P,2019ApJ...881..125D,2019ApJ...884...15S}.

In this work we describe the multi-wavelength emission from \PKS~and its possibly associated neutrino emission.  While earlier models frequently assume that the gamma rays are produced by neutral pion decays accompanying the neutrino production \citep{2016NatPh..12..807K}, the observation of neutrinos from TXS 0506+056 has taught us that observational constraints in the X-ray band can limit the hadronic contribution. In that case, the neutrino flux has been shown to be explained by a hybrid leptohadronic model \citep[e.g.][]{Gao:2018mnu, Keivani:2018rnh, Oikonomou:2019djc}. We apply this model to \PKS where the neutrino-producing hadronic processes only contribute to X-ray and possibly TeV gamma rays, where the rest of the multi-wavelength spectrum are produced leptonically. We also examine another possibility, where the GeV gamma rays are explained by proton synchrotron emission \citep[see e.g.][]{Diltz:2015kha, Cerruti:2018tmc}. Since \PKS falls in the blazar sub-class of flat-spectrum radio quasars (FSRQs), the GeV emission can originate from inverse-Compton scattering of radiation fields external to the jet \citep{Rodrigues:2018tku}. This effect is included in the models explored in this work.

\section{Methods} \label{sec:methods}

In this section, we first explain the process for identifying the different relevant epochs in terms of gamma-ray emission. We then describe the radiation model, including the treatment of external radiation fields. Finally, we discuss the methods used to search the parameter space of the source.

\subsection{Analysis of the multi-wavelength behaviour}
\label{sec:methods_lightcurve}
The majority of the observed gamma-ray blazars exhibit a highly variable behavior on timescales from minutes to years \citep{2019ApJ...877...39M}. Measurements of these characteristic timescales and the spectral properties of the sources during these bright flaring states can provide useful insights about the emission region and the mechanisms behind the production of the observed gamma rays.

\begin{figure*}[htbp]
    \includegraphics[width=\linewidth,trim=43 0 85 0, clip]{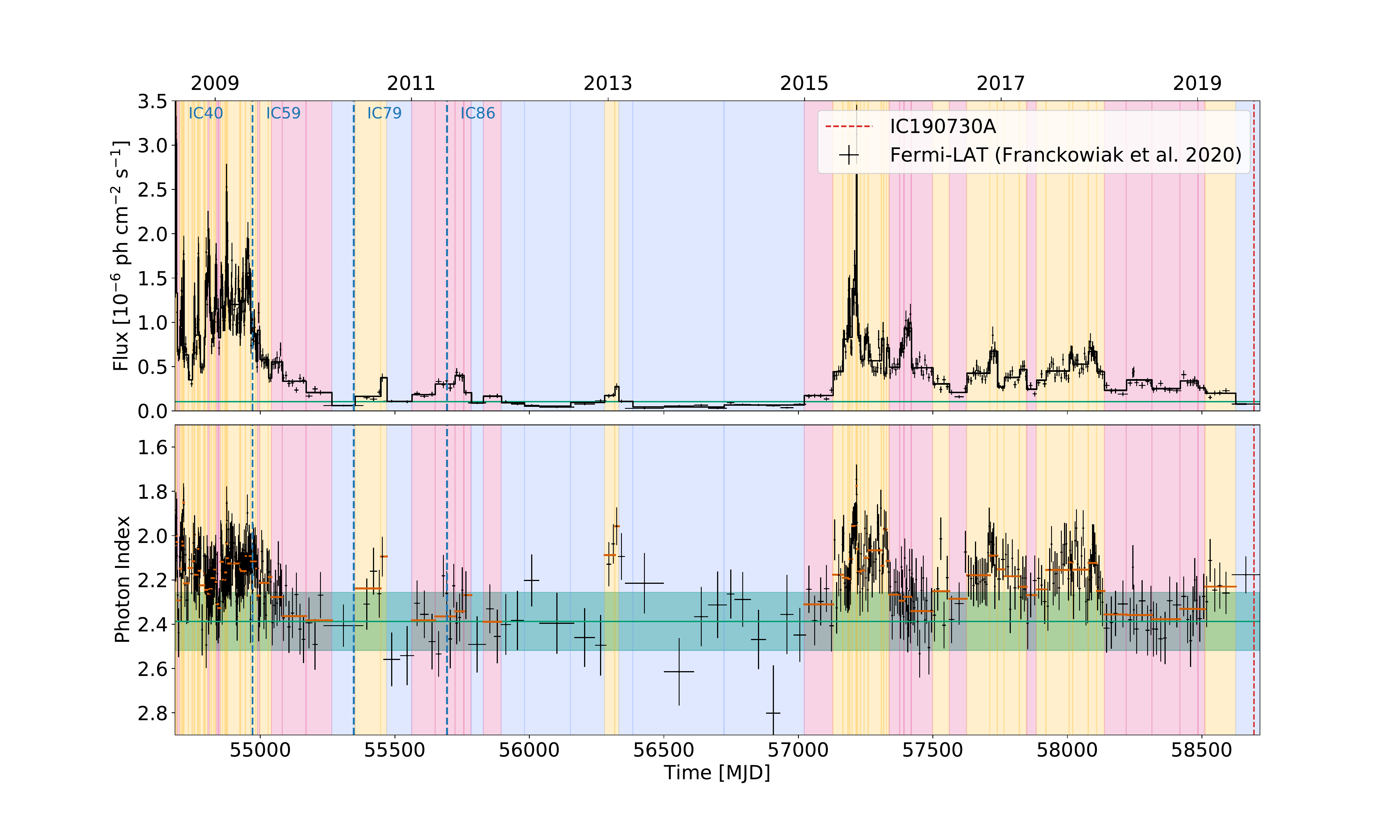}
    \caption{\textit{Top:} eleven-year \textit{Fermi}-LAT lightcurve of \PKS \citep{Franckowiak:2020qrq}, divided for the purposes of this analysis into three characteristic states: \textit{quiescent} (blue), flaring with a hard gamma-ray spectrum (\textit{hard flares}, yellow) and flaring with a soft gamma-ray spectrum (\textit{soft flares}, pink). The green line shows the average flux of 3$\times$10$^{-7}$ ph cm$^{-2}$ s$^{-1}$. \textit{Bottom}: \textit{Fermi}-LAT spectral index across the same 11 year time period. The green line shows the average spectral index ($\bar\Gamma = 2.31$), from where we draw the distinction between hard ($\Gamma<\bar\Gamma$) and soft  ($\Gamma>\bar\Gamma$) flares, and the green band is its respective 1$\sigma$ uncertainty. The red lines show the average spectral index in the time window of each flare.}
    \label{fig:slots}
\end{figure*}

We define a simple method to distinguish flaring states from the quiescent state of the source. To that end we utilize the gamma-ray lightcurve for 11 years of observation of \PKS provided by \cite{Franckowiak:2020qrq}, as shown in the top panel of \Fig~\ref{fig:slots}, and the corresponding \textit{Fermi}-LAT spectral indices, shown in the bottom panel. While there is no standard method to define a quiescent state in a blazar lightcurve, we consider the average of the measured fluxes in the low-activity period from 2010 February to 2014 December weighted by the time duration of each bin as a proxy for the quiescent state. In order to identify the brightest flaring periods for the source, we use the Bayesian blocks representation of the lightcurve \citep{2013ApJ...764..167S} and we define all the Bayesian blocks with a flux level higher than the calculated average as flaring states of the source.

We define three different multi-wavelength states for \PKS based on the gamma-ray data, comparing the average index in each Bayesian Block with the average index and error from the quiescent proxy:  \begin{enumerate*}[label=(\roman*)]
\item a \textit{quiescent state}, with low gamma-ray flux
\label{item:quiescent},
\item a gamma-ray flaring state with hard spectral index compared to the average observed value of $\bar\Gamma$ = (2.39 $\pm$ 0.13) 
\textit{(hard flares)} and \label{item:high_hard}
\item a gamma-ray flaring state with spectral index compatible to the one observed during the quiescent proxy period, within the average uncertainty band.  
\textit{(soft flares)}\label{item:high_soft}.
\end{enumerate*}

In general, we find a good correlation between the gamma-ray and optical fluxes in \cite{Franckowiak:2020qrq}. The X-ray sampling is poor and the variability observed in radio surveys is slow, so observations in these bands are not suitable for further refining the definition of the different states.

The three activity states are highlighted in different colors in \Fig~\ref{fig:slots}. The 11 year period is characterized by a quiescent state, with a total duration of about 3.8 years (blue shaded areas). The detection of IceCube-190730A falls into this state, as indicated by the dashed red line. The total duration of the other states are 3.7 years for the hard flaring state and 3.5 years for the soft flaring state, highlighted in \Fig~\ref{fig:slots} with yellow and pink shaded areas, respectively.

In \tabl{flares} we report the three observation periods that will be used to represent each of the three activity states of the source later in this work, when the numerical radiation models are used to describe the multi-wavelength emission. These simultaneous multi-wavelength data sets were adopted from \cite{Franckowiak:2020qrq}.

\begin{table}[htbp]
    \centering
    \caption{Information about the three activity states of \PKS identified in \Fig~\ref{fig:slots}. The representative datasets refer to multi-wavelength observations from each of the three activity states, which are used in this work to constrain the source models. The quiescent state observations are coincident with the neutrino detection. The three datasets were compiled by \citet{Franckowiak:2020qrq}: the quiescent state data can be found in the bottom right panel of Fig. 4 of that work, and the data from the two flaring states in the bottom right panel of the same figure.}.

    \begin{tabular}{ccc}
         \toprule
         State      &  Total duration [yr] & Representative dataset \\
         \midrule
         Quiescent  &  3.8 &                 MJD 58664-58724\\
         Hard Flare &  3.7 &                 MJD 57210-57219\\
         Soft Flare &  3.5 &                 MJD 58107-58125\\
        \bottomrule
    \end{tabular}
    \label{tab:flares}
\end{table}

\subsection{Numerical radiation model}

For each of the three states, we numerically model the multi-wavelength and neutrino emission of \PKS using the time-dependent simulation code AM$^3$ \citep{Gao:2016uld} which solves the system of coupled differential equations describing the transport of all relevant particles interacting in the blazar jet. Non-thermal electrons and protons are assumed to be accelerated and subsequently injected into a single radiative zone in the jet. This zone is modeled as a blob that is spherical in its co-moving frame and spatially homogeneous. Although there may be multiple emitting regions in the jet of \PKS, in the models explored here this single zone is responsible for the multi-wavelength emission between the optical and the gamma ray regimes, as well as neutrino production.  

The comoving radius of the blob is denoted as $R^{\prime}_{\mathrm{b}}$ (quantities in the blob's comoving frame are primed). We assume the magnetic field is randomly oriented within the blob with a homogeneous strength $B^{\prime}$. The blob is moving at relativistic speed relative to the supermassive black hole with a Lorentz factor of $\Gamma_{b}$ and we assume the jet is observed at an angle $\theta_{\mathrm{obs}}=1/\Gamma_{b}$ relative to its axis, resulting in a Doppler factor of $\delta_{D}=\Gamma_{b}$. The distance of the blob to the supermassive black hole is parameterized by the dissipation radius, $R_\mathrm{diss}$, and its impact on the model is discussed in \Sec~\ref{sec:external}.

Electrons are assumed to be accelerated to a simple power-law distribution, $dN_{e}/d\gamma_e\propto \gamma^{-p_{e}}$, from a minimum to a maximum Lorentz factor,  $\gamma_e^{\mathrm{min}}$ and $\gamma_e^{\mathrm{max}}$. For protons we also test whether the observations can be better explained with a break on the spectrum at a Lorentz factor $\gamma_p^{\mathrm{min}} < \gamma_p^{\mathrm{break}} < \gamma_p^{\mathrm{max}}$, where the spectral index changes from $p_p^\mathrm{low}$ to $p_p^\mathrm{high} > p_p^\mathrm{low}$. The normalization of these acceleration spectra is quantified by means of the total (energy-integrated) luminosity deposited into non-thermal electrons, $L^\prime_e$, and protons, $L^\prime_p$.

\subsection{External radiation fields}
\label{sec:external}
Being an FSRQ, \PKS possesses a broad line region (BLR) surrounding the accretion disk, which reprocesses and partially isotropizes the emission from the powerful accretion disk surrounding the black hole. The emission from the disk can be observed as a thermal bump in the optical/ultraviolet (UV) regime ($\sim5~\mathrm{eV}$) during the quiescent state of the blazar \citep{Franckowiak:2020qrq}, see also \Sec~\ref{sec:results}. The observed UV flux translates to a luminosity of $L_\mathrm{disk}=2.6\times10^{46}~\mathrm{erg/s}$. This is consistent with the value $2\times10^{46}~\mathrm{erg/s}$, computed using broad Mg~{\sc ii} and C~{\sc iv} emission line luminosities, as reported by \citet{2011ApJS..194...45S} and adopting the scaling factors proposed by \citet{1991ApJ...373..465F}.

Following \citet{Ghisellini:2009wa}, we assume the BLR to be a thin shell located at a radius $R_\mathrm{BLR}=5\times10^{17}L_\mathrm{disk, 46}^{1/2}~\mathrm{cm}$ around the supermassive black hole. We further assume that the BLR re-processes about 10\% of the power output by the accretion disk~\citep{Greene:2005nj}, re-emitting it isotropically in the rest frame of the black hole\footnote{Most of the disk radiation processed by the BLR is in fact re-emitted as atomic lines and not a thermal continuum~\citep{Greene:2005nj}. However, the results of this study are not affected by this distinction, since most of the atomic emission will lie on a similar frequency range as the thermal disk emission (the Ly $\alpha$ line has an energy of 10~eV for hydrogen).}.
Inside the volume surrounded by the BLR, the energy density of this isotropic field is constant and proportional to $L_\mathrm{disk}R_\mathrm{BLR}^{-2}$. In the rest frame of the jet, this energy density receives a relativistic boost given by $\Gamma_\mathrm{b}^2$, and the photon frequencies are Doppler-shifted by a factor $\Gamma_\mathrm{b}$. 

Outside the BLR, the energy density of the external fields declines with distance to the black hole according to \eqs~19 and 20 from \citet{Ghisellini:2009wa}. Therefore, in the case where the blob lies outside the BLR ($R_\mathrm{diss}>R_\mathrm{BLR}$), the energy density of the external fields seen by the particles in the blob depends inversely on the dissipation radius  $R_\mathrm{diss}$. 

Additionally to the disk radiation reprocessed by the BLR, we also consider thermal infrared emission from a dusty torus surrounding the disk. However, because of the larger volume spanned by this emission, the corresponding photon energy density is negligible compared to the BLR emission in all the cases considered. The inclusion of a sub-dominant component up to X-rays from the black hole corona would also not affect the results in any of the cases modeled.

\subsection{Parameter search}

We have searched the parameter space of the source in two distinct regimes, distinguished primarily by the strength of the magnetic field in the jet: in the \textit{leptohadronic model} we admit values of $B^\prime\leq1~\mathrm{G}$, while in the \textit{proton synchroton model} these values are higher, $B^\prime\geq10~\mathrm{G}$. In this regime, proton synchrotron emission can contribute to the observed high-energy fluxes, while in the leptohadronic model protons will only contribute significantly through photo-hadronic interactions.

The parameter space was scanned using a genetic algorithm similar to that used by \citet{Rodrigues:2018tku}, with the purpose of minimizing the $\chi^2$ value of the predicted multi-wavelength fluxes compared to data. Additionally to using an efficient search algorithm, the method also involves the simulation of a large number of parameter sets, in the order of $10^6$  for each activity state.

\section{Results} \label{sec:results}

Using the numerical models described in the previous section, we have calculated the multi-wavelength and neutrino spectra for \PKS during the different epochs. We have then compared the neutrino results to the statistical expectations based on the IceCube sensitivity.

\subsection{Multi-messenger emission}

\begin{figure*}
    \includegraphics[width=0.5\linewidth]{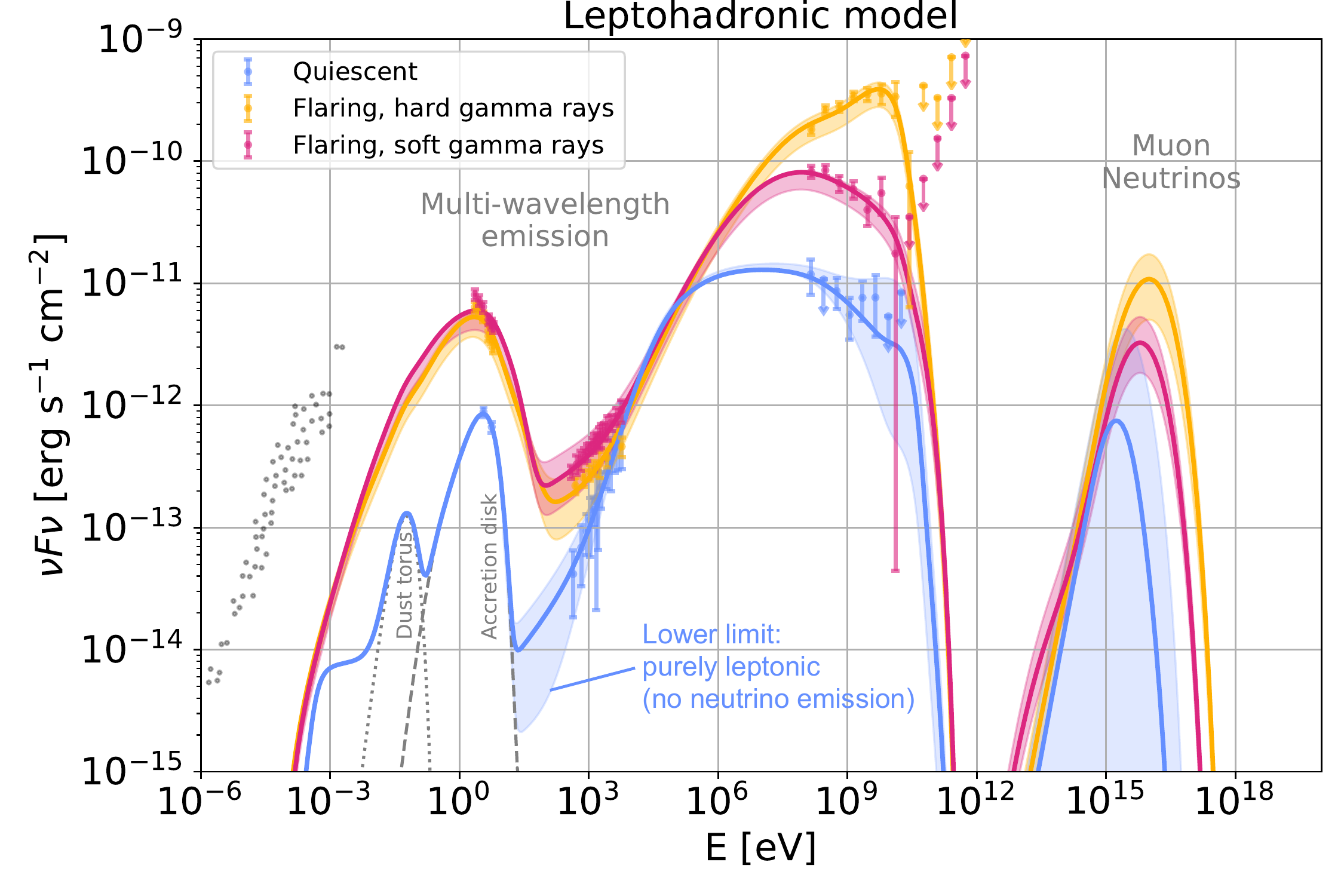}
    \includegraphics[width=0.5\linewidth]{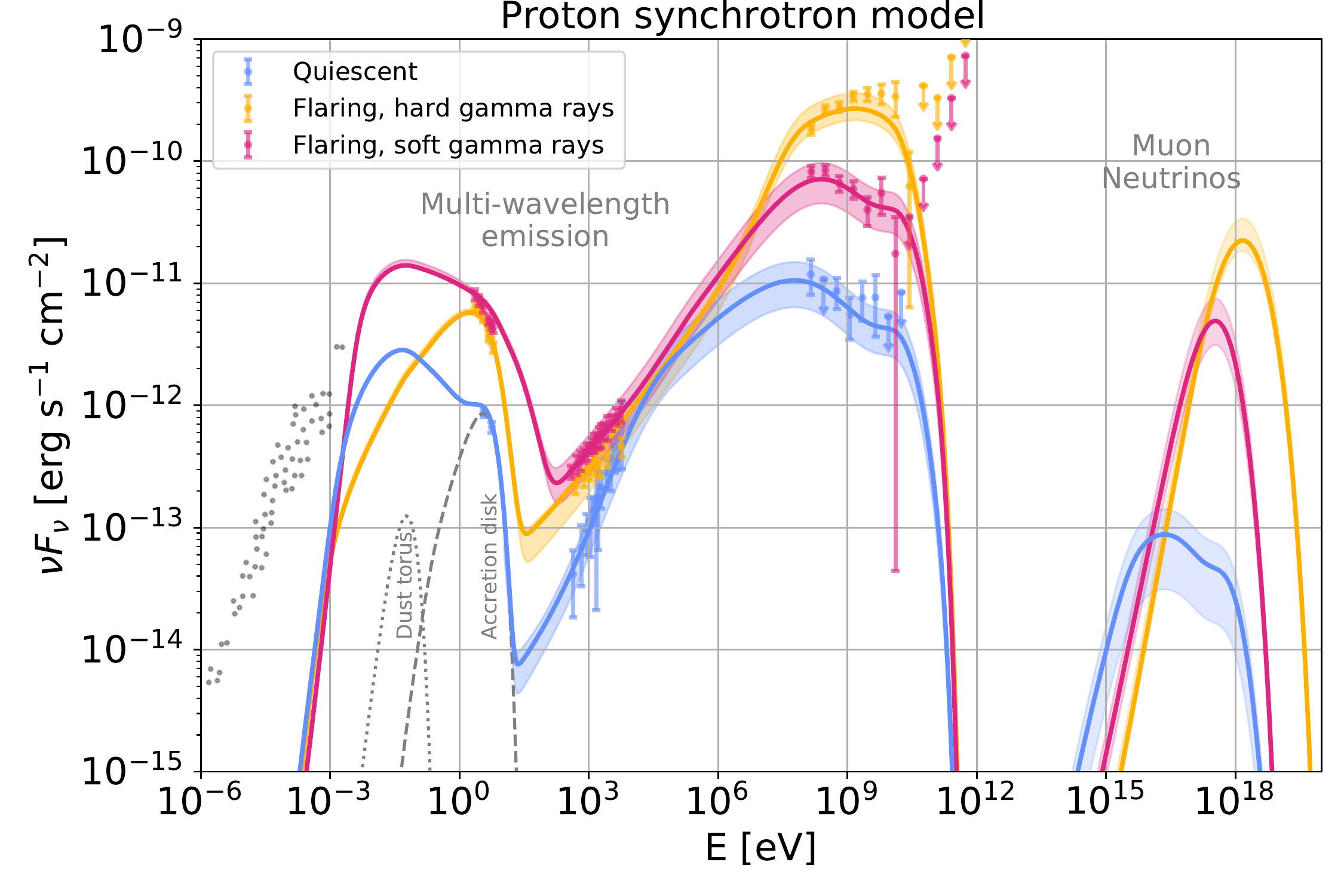}
    \caption{The colored curves show the predicted multi-wavelength fluxes and all-flavor neutrino spectra from \PKS obtained with the leptohadronic model \textit{(left)} and the proton synchrotron model \textit{(right)} under three different parameter sets, indicated in \Tab~\ref{tab:parameters}. The shaded areas correspond to the uncertainty in the non-thermal proton power, also indicated in \Tab~\ref{tab:parameters}. The colored data points represent multi-wavelength flux observations during each of the three states  (see \Fig~\ref{fig:slots} and \Tab~\ref{tab:flares}). The gray points show archival radio data from the source. The interaction zone responsible for the optical/UV, X-ray, gamma-ray and neutrino emission is too compact to produce this radio emission (due to strong synchrotron self-absorption). The radio flux must therefore originate in synchrotron emission from a larger region of the jet, and is therefore uncorrelated with neutrino production in these models.}
    \label{fig:seds}
\end{figure*}

The emitted fluxes during the three activity states of \PKS are shown in \Fig~\ref{fig:seds}, as predicted by the leptohadronic and the proton synchrotron models (left and right panels, respectively). 
The fluxes shown as colored data points correspond to three well-covered multi-wavelength observations analyzed in \cite{Franckowiak:2020qrq} that are representative of the different states considered for the source in this work: in blue we represent observations during the quiescent state, in yellow during hard flares, and in pink for soft flares. The data points shown in the figure were the ones used to fit the model in each of the three epochs. The gray points in the radio band correspond to historical radio data.
The neutrino-emitting region in the jet (i.e. the blob) is necessarily too compact to explain the archival radio observations from the source. This is because the high radiation density necessary for hadronic processes leads to efficient synchrotron self-absorption at low frequencies, which limits the outgoing radio flux. Therefore, we assume that the radio observations originate from electrons radiating in a more extended region of the jet.

Both the multi-wavelength and neutrino fluxes have a best-fit result, represented by the colored curves, and an uncertainty band. The best-fit parameters are listed in  \Tab~\ref{tab:parameters}. The uncertainty band is obtained by varying the power in accelerated protons (while keeping all other parameters constant) until the flux (in either X-rays or gamma rays) deviates from the best-fit model by $\pm40\%$. This variation corresponds roughly to the $1~\sigma$ spread we find in the individual flux bins throughout each of the three activity states.
This band is intended to reflect the variability of the high-energy emission observed in the duration of a given activity state, which cannot be fully represented by a single set of simultaneous data. This variability in the data leads to an uncertainty in the neutrino flux emitted by the source throughout each of the three states, which is not necessarily constant; the uncertainty bands allow us to assess this uncertainty when evaluating the predicted number of events in IceCube.
At the same time, we require the models to describe well the simultaneous multi-wavelength data that is available (namely the three data sets shown in \Fig~\ref{fig:seds}). We therefore use these three data sets to test the goodness of fit of the models, which are reported in \Tab~\ref{tab:parameters} in the form of a $\chi^2$ value per degree of freedom.

In both models, the observed UV fluxes result from electron synchrotron emission during the flaring states, and from an exposed accretion disk during the quiescent state. The differences between the models impact primarily the high-energy emission: in the right-hand panel of \Fig~\ref{fig:seds}, proton synchrotron dominates the emitted gamma-ray flux below 1~GeV and, in the quiescent state, also the X-rays. As we can see in \Tab~\ref{tab:parameters}, the minimum proton Lorentz factor required in this model can reach values up to $\gamma_p^\mathrm{min}\sim10^8$, which is necessary to ensure that proton synchrotron emission is not significant below X-ray frequencies.

On the contrary, in the leptohadronic model (left panel), gamma-ray emission is mostly dominated by external Compton scattering. This is possible due to the location of the blob near the perimeter of the BLR ($R_\mathrm{diss}\sim R_\mathrm{BLR}$, as listed in \Tab~\ref{tab:parameters}), while in the proton synchrotron model the blob lies well outside the BLR and thus external fields do not play a significant role. On the other hand, the photons from hadronic processes explain the X-ray observations, especially during the flaring states. 

As we can see by the blue band in the left-hand panel of \Fig~\ref{fig:seds}, the quiescent state can be fit within a range of proton injection luminosities, which lead to different levels of neutrino emission. The best fit, represented by the blue curves, has a hadronic component, which is responsible for neutrino emission. When this hadronic component is completely removed, we obtain the lower limit of the blue band, and there is no neutrino emission (the blue neutrino band extends down to zero). In this purely leptonic limit, the simultaneous data shown in blue is not fit as well in X-rays and gamma rays above 1~GeV. However, as mentioned earlier, all results within the colored bands lie within the $1~\sigma$ spread in the fluxes observed during the quiescent state in the 11 year lightcurve. Therefore, the quiescent state of the source is in general compatible with a purely leptonic scenario.

Contrary to the quiescent state, our parameter search has revealed that the flaring states are not easily explained by a purely leptonic scenario. The relatively bright and soft X-ray spectrum (see pink and yellow data points) must harden around MeV energies in order to explain the high gamma-ray fluxes, especially during the hard gamma-ray flares. As explained in detail below, in both the proton synchrotron and leptohadronic models these X-rays originate in cascades initiated by high-energy hadronic photons, which provide a necessary component to bridge the two humps of the emission spectrum.

\begin{figure*}
    \includegraphics[width=0.5\linewidth]{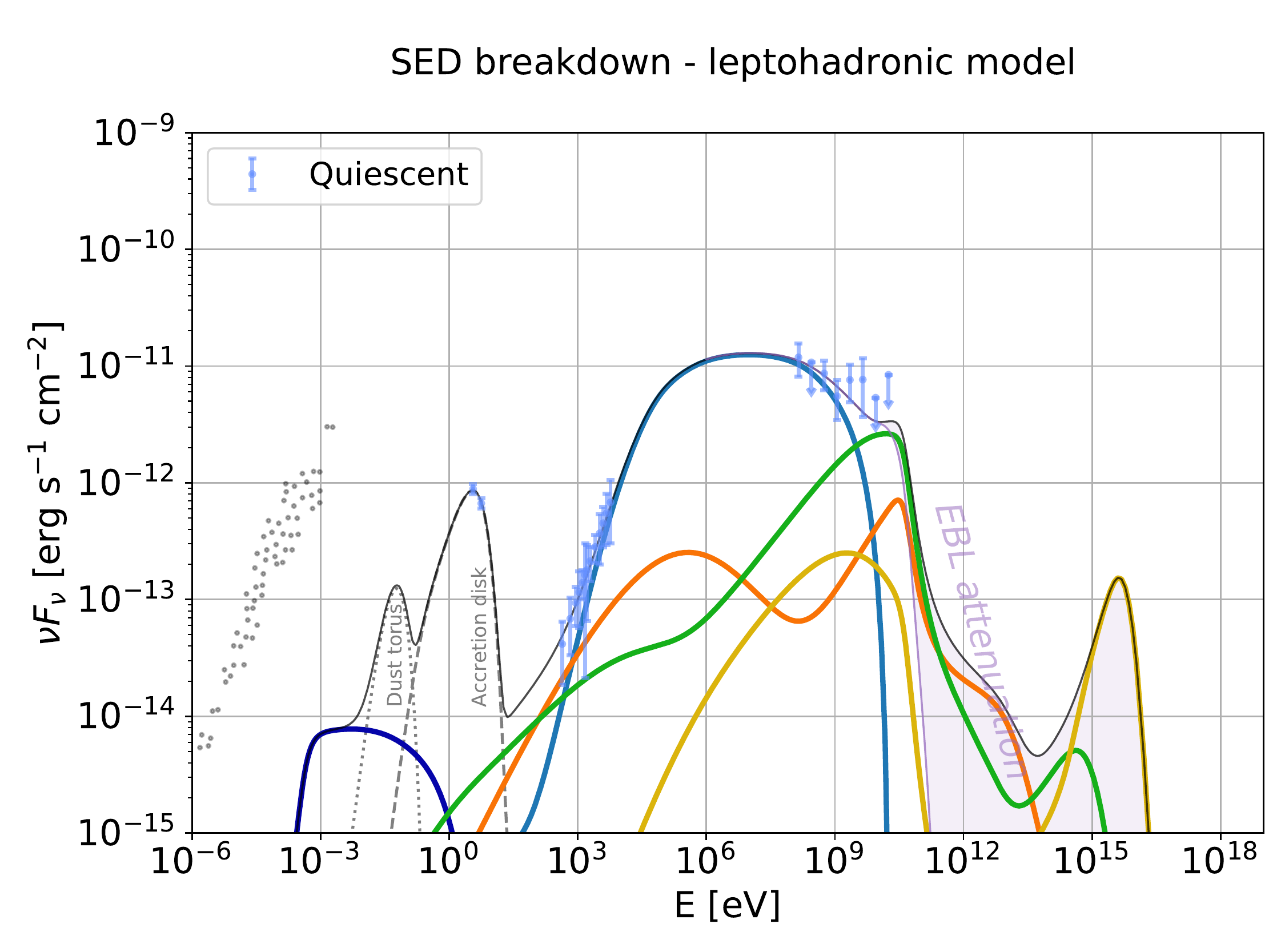}\includegraphics[width=.5\linewidth]{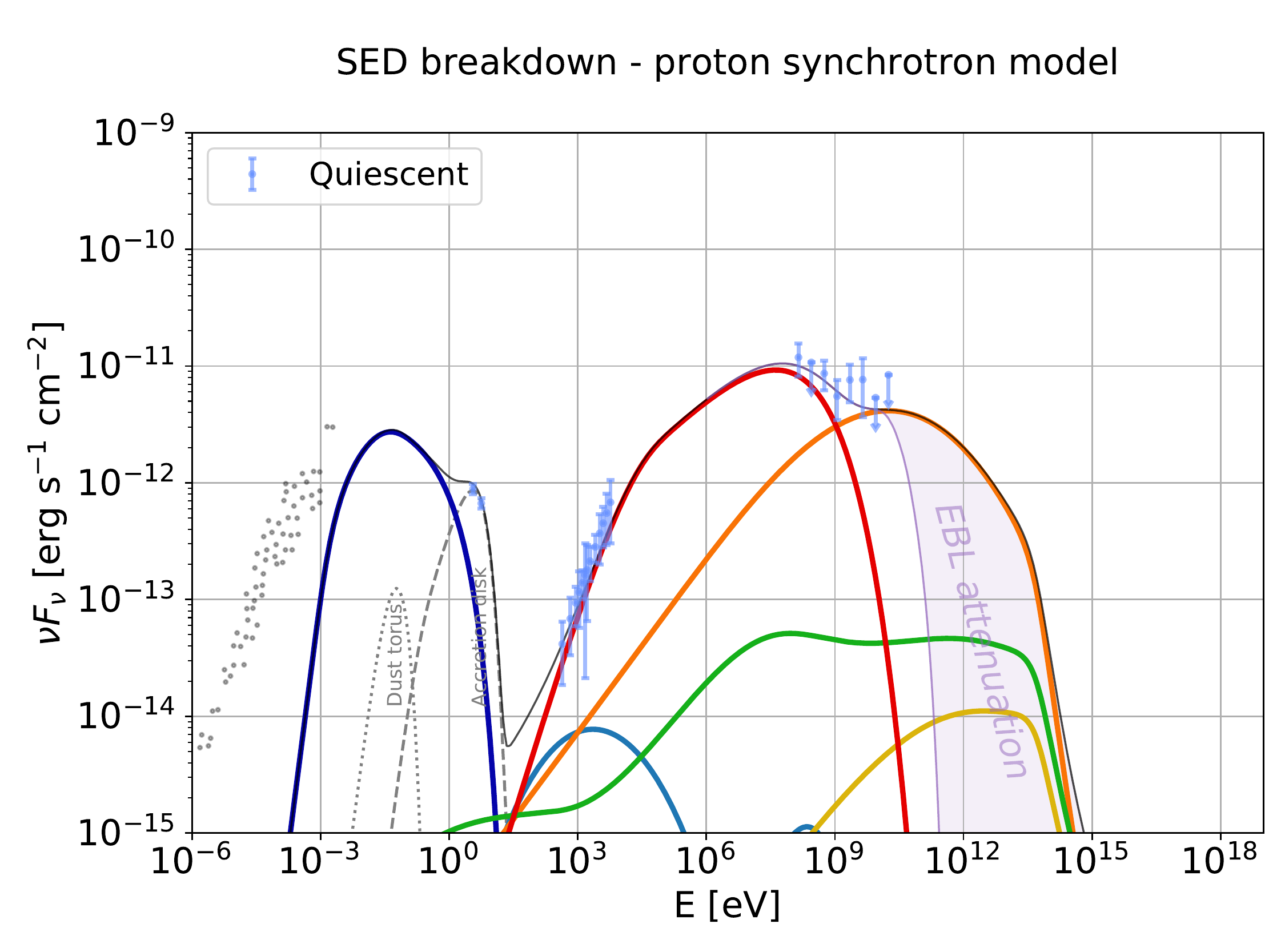}
    \includegraphics[width=0.5\linewidth]{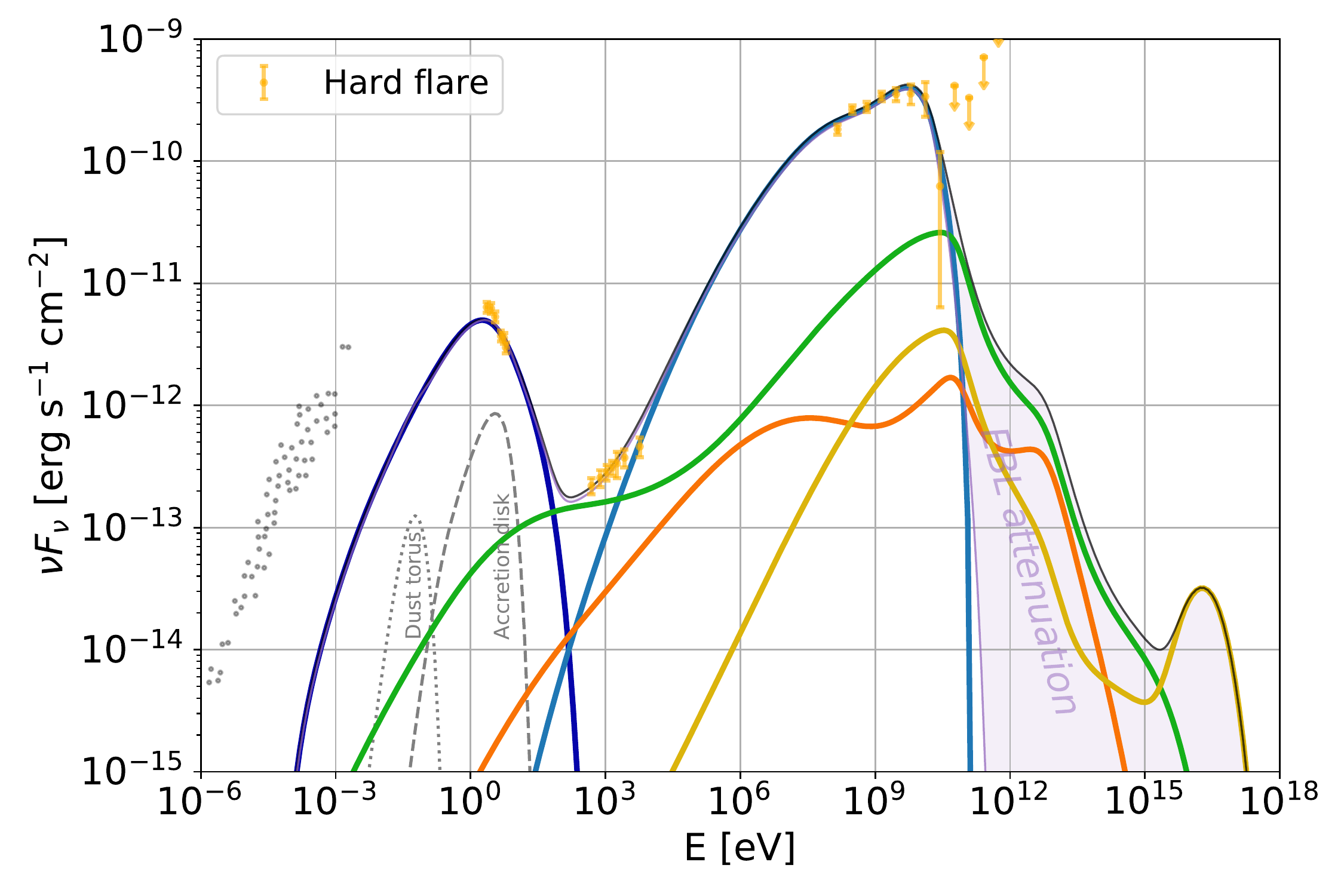}\includegraphics[width=.5\linewidth]{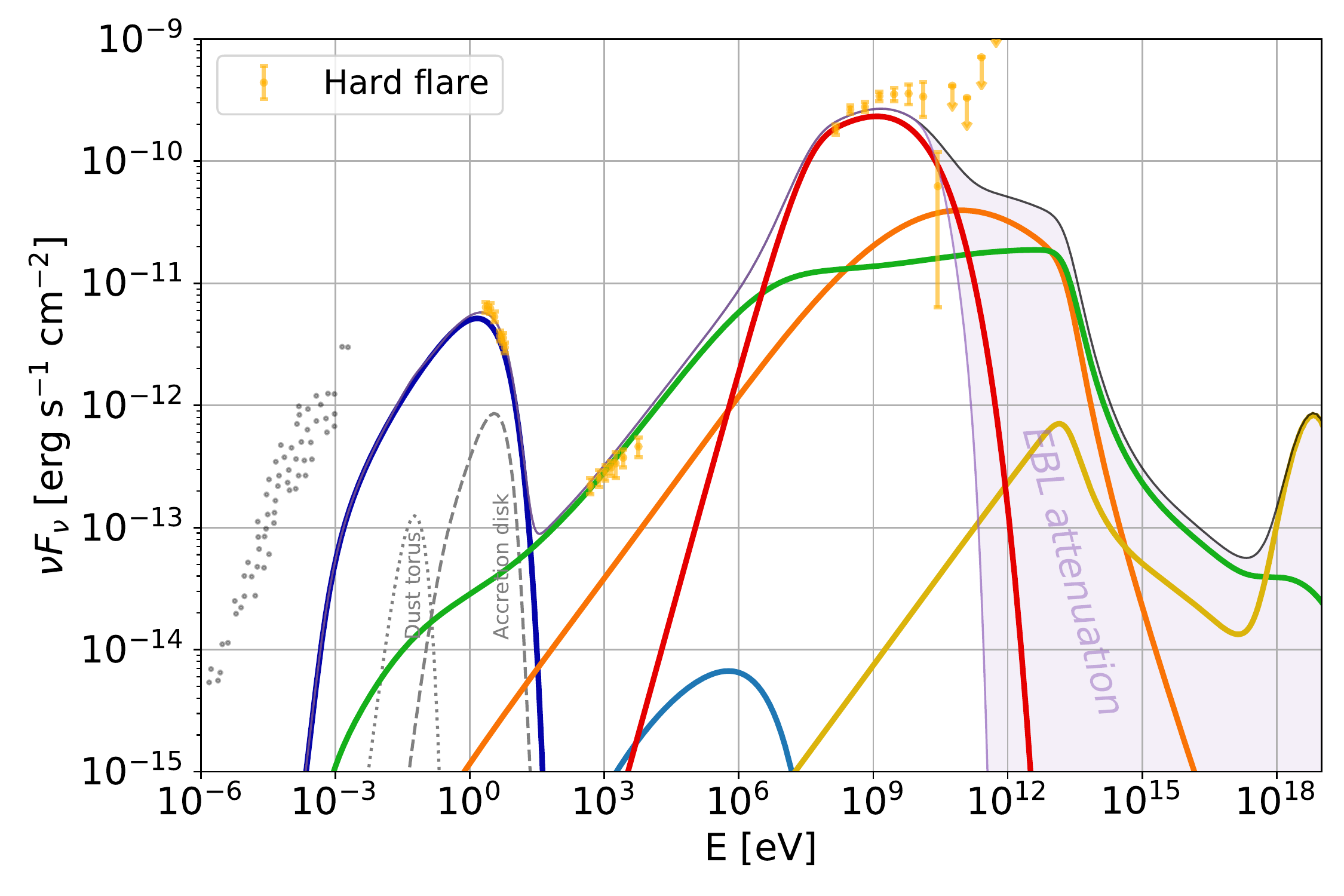}
    \includegraphics[width=0.5\linewidth]{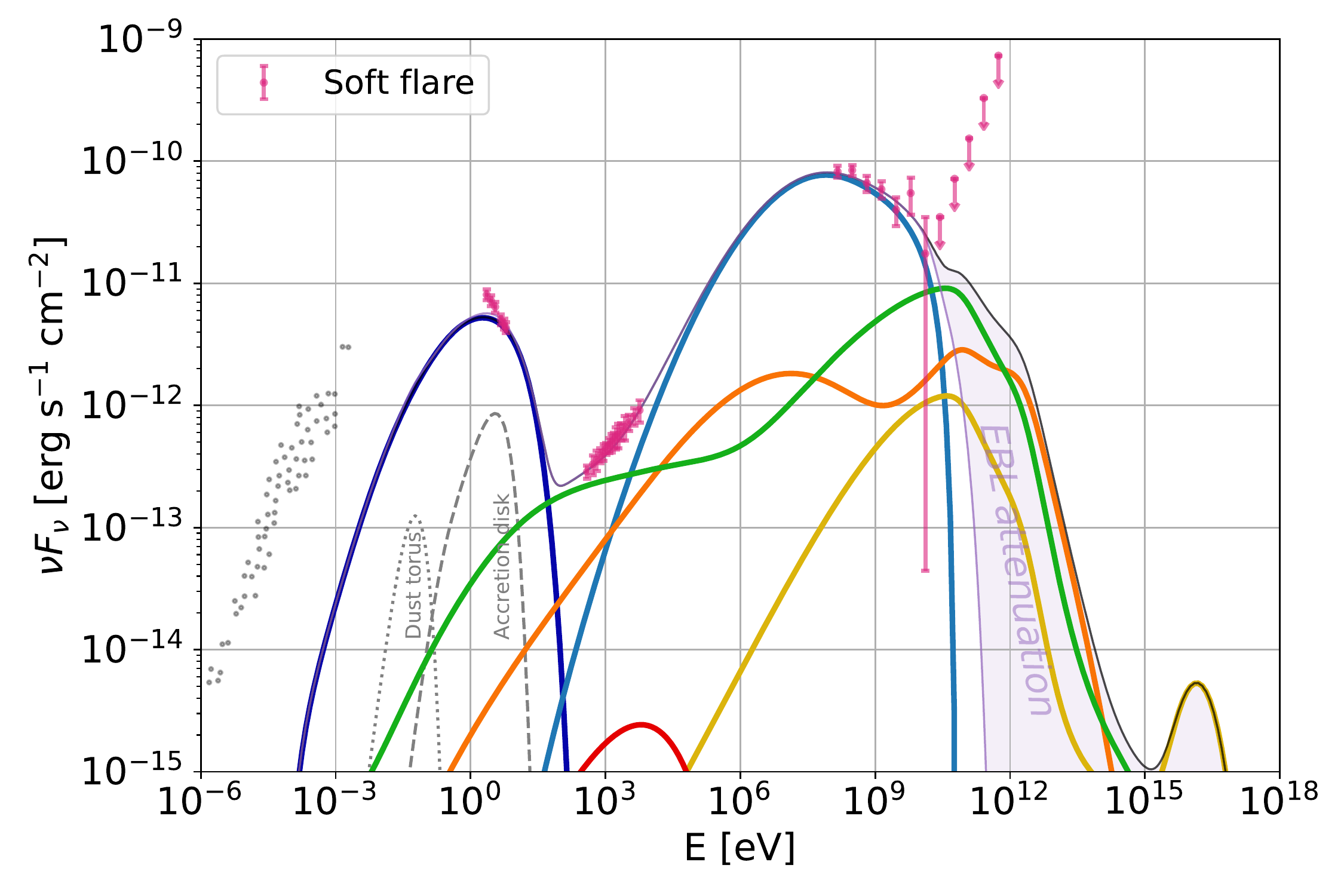}\includegraphics[width=.5\linewidth]{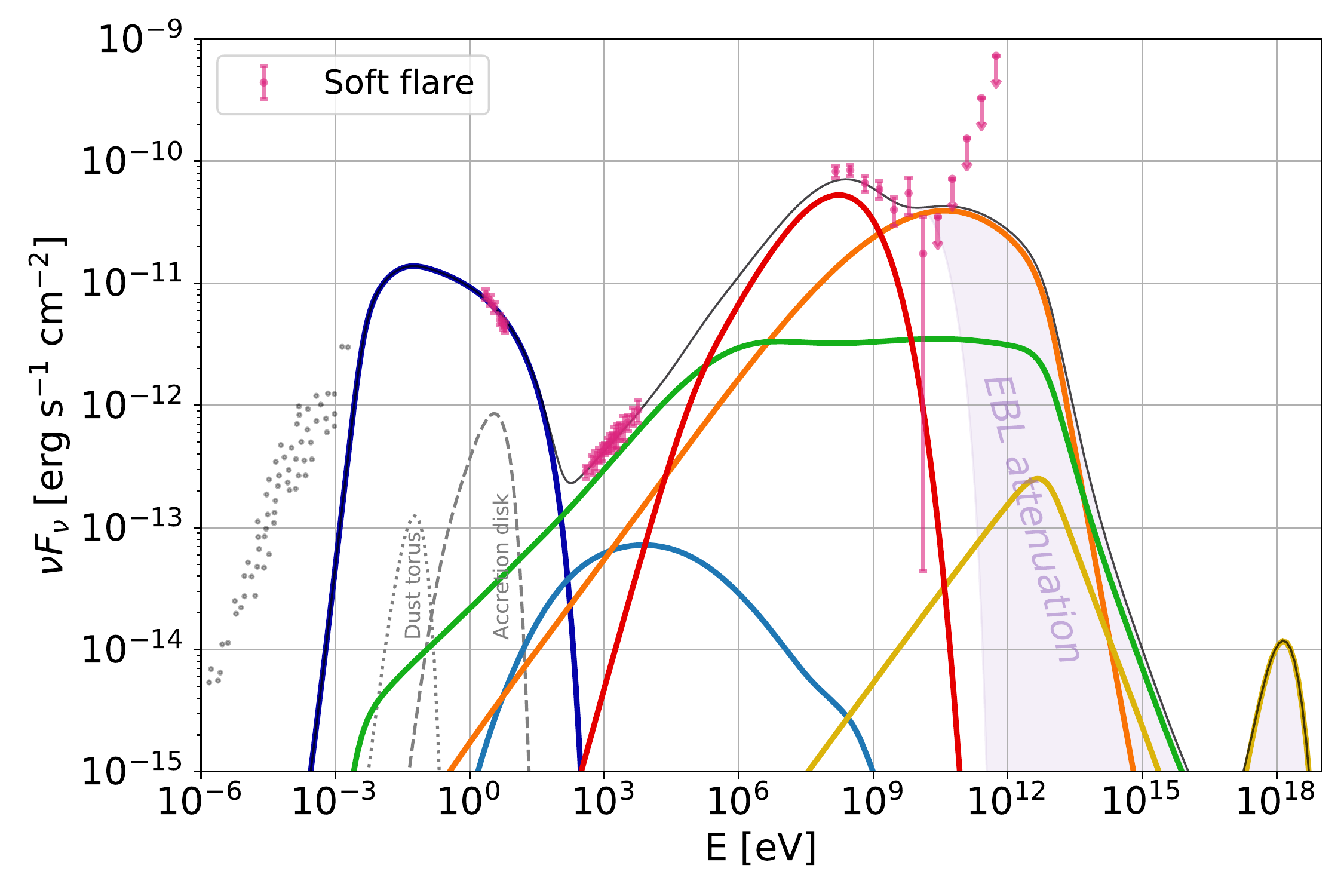}
    \includegraphics[width=\linewidth]{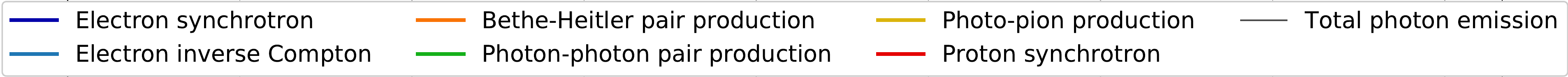}
    \caption{Breakdown of the spectrum during the different states into the different radiative processes for the leptohadronic model \textit{(left)} and the proton synchrotron model \textit{(right)}.}
    \label{fig:breakdown}
\end{figure*}

In order to help understand the details of the two models, in \Fig~\ref{fig:breakdown} we break down the multi-wavelength fluxes shown in \Fig~\ref{fig:seds} into their different radiative components. In the three left panels, we show the processes responsible for the emission in the leptohadronic model. As mentioned earlier, gamma-ray fluxes are dominated by Compton scattering (light blue curve) of the external thermal fields. Additionally, the accelerated protons emit photons through photo-pion production (yellow) and Bethe-Heitler pairs, which in turn radiate through synchrotron and inverse Compton (orange). When these high-energy photons annihilate with lower-energy target photons, an electromagnetic cascade is created in the jet, whose emission is shown in green. Above 100 GeV, the emitted radiation is strongly attenuated by EBL interactions, as represented by the purple band.

Considering only the leptonic emission, we would have necessarily a deep gap between UV and X-rays, and the inverse Compton emission provides a hard spectrum between X-rays and gamma rays. In the quiescent state  (upper left panel), this hard inverse Compton spectrum can explain the X-ray observations above 1~keV, while the photons from cascades and Bethe-Heitler emission contribute to the soft X-rays. On the contrary, in the flaring states (middle and lower panels), the observed X-ray flux is softer. The cascade emission is therefore necessary in this model to explain observations in this energy range. This seems to provide some evidence of proton interactions in the source solely from the perspective of the multi-wavelength behaviour of the source.

Additionally to X-rays, the cascades from hadronic photons also contribute significantly to the gamma-ray flux above 1 GeV in the quiescent state. As shown previously in \Fig~\ref{fig:seds}, when the hadronic component is removed completely, the two \textit{Fermi}-LAT data points at the highest energies are not explained, leading to a worse fit. On the other hand, gamma rays below 1 GeV are independent of this hadronic contribution.

In the right panels of \Fig~\ref{fig:breakdown} we show a breakdown of the emission in the proton synchrotron model. In the quiescent state the X-ray fluxes are explained by proton synchrotron emission, as well as gamma rays up to 100 MeV. Above this energy, the spectrum is dominated by emission from Bethe-Heitler pairs. In the flaring states, when the observed X-ray spectrum is softer, it is the cascade emission that dominates that energy range, such as in the leptohadronic model.

Regarding neutrino emission, the proton synchrotron model predicts a high peak energy of around 10~PeV to 1~EeV, while in the leptohadronic model the neutrinos peak around 1-10~PeV.
The neutrino energies are determined by the maximum proton energy, which is constrained by observations in both models. In the proton synchrotron model, it is constrained by gamma-ray observations, since these are explained by proton synchrotron emission. In the leptonic model, the maximum protons energy is constrained mainly by the X-ray fluxes, since the extent of the electromagnetic cascade depends on the energy of the interacting protons.

Importantly, the leptohadronic model predicts a quiescent-state neutrino flux a factor 10 higher compared to proton synchrotron. This is due to the high density of radiation from the accretion disk whose energy is boosted in the jet rest frame up to the photo-pion production threshold. In the proton synchrotron model, on the other hand, the main target for photo-pion production are non-thermal photons. During the flaring states, there is a high density of UV photons from electron synchrotron, thus enhancing neutrino emission. During the quiescent state, this non-thermal emission is dim, and the neutrino production is therefore low. The sharp dip in the photon spectrum between the cutoff of the electron synchrotron and the onset of the proton synchrotron leads to a double hump in the neutrino spectrum that can be seen in \Fig~\ref{fig:seds}. In the leptohadronic model there are no such structures in the neutrino spectrum because it is the external photons that provide the main target for photo-pion production.

During flares, the proton synchrotron model predicts a higher-energy flux of neutrinos, but since the spectrum is harder and narrower than in the leptohadronic case, the corresponding total number of neutrinos is in fact lower. This will reflect on the predicted number of IceCube events, as discussed in the next section.

\begin{table*}
    
    \caption{Parameter values underlying the results of the leptohadronic and proton synchrotron models, for each of the states identified in \Fig~\ref{fig:slots}. Primed quantities refer to the rest frame of the jet. The ranges in the values of the proton luminosity correspond to the uncertainties of the model, resulting in the shaded regions in \Fig~\ref{fig:seds}. We also report the reduced $\chi^2$ values for the multi-wavelength SEDs predicted by each model, describing the goodness of fit. In the two bottom rows, we list the predicted number of neutrino events per year in IceCube, as well as the total expected number of events integrated over the eleven-year period. The yearly rates correspond to the IC86 configuration of the IceCube detector, since they were calculated using the effective area of that configuration. For the total number of events, we take into account the different detector configurations over the years, as depicted by the dashed vertical lines in \Fig~\ref{fig:slots}.}
    
    \centering
    \begin{tabular}{ccccccc}
         \toprule
         Model & \multicolumn{3}{c}{Leptohadronic} & \multicolumn{3}{c}{Proton Synchrotron}\\
         State      &    Quiescent    &    Hard Flare    &    Soft Flare        &    Quiescent    &    Hard Flare    &    Soft Flare\\
         \midrule
        $R^\prime_\mathrm{b}$ [cm, log]         & 16.0& 15.9& 15.9 & 16.0& 16.0& 16.0 \\
        $B ^\prime$ [G]                         & 0.3 & 0.3 & 0.6 & 10.0& 12.6& 15.8 \\
        Bulk Lorentz factor $\Gamma_\mathrm{b}$ & 27.6& 28.7& 26.2 & 40.0& 49.2& 42.6 \\
        $R_\mathrm{diss}/R_\mathrm{BLR}$        & $\leq1.0$ \phantom{--} 
                                                & 1.2
                                                & 1.4
                                                &$\geq2.6$\phantom{--}
                                                & $\geq3.6$\phantom{--} 
                                                & $\geq2.6$\phantom{--} \\
        $L^\prime_e$  [erg s$^{-1}$, log]       & 43.5& 44.6& 44.2 & 42.0& 41.6& 42.6 \\
        $L^\prime_p$ [erg s$^{-1}$, log]        & \phantom{-} $\leq45.7 ^{+0.8}$ 
                                                & \phantom{----} $46.5 _{-0.2} ^{+0.2}$ 
                                                & \phantom{----} $46.9 _{-0.1} ^{+0.2}$ 
                                                & \phantom{----} $46.4 _{-0.3} ^{+0.4}$ 
                                                & \phantom{----} $46.2 _{-0.2} ^{+0.0}$ 
                                                & \phantom{----} $46.0 _{-0.2} ^{+0.1}$\\
        $\gamma_e^{\prime\mathrm{min}}$, log  & 1.0 & 3.8 & 3.3 & 2.0 & 3.0 & 1.9 \\
        $\gamma_e^{\prime\mathrm{max}}$, log  & 3.7 & 4.5 & 4.2 & 3.0 & 3.1 & 3.5\\
        $p_e$                                 & 2.1 & 3.6 & 1.2 & 2.1 & 3.5 & 2.1 \\
        $\gamma_p^{\prime\mathrm{min}}$, log  & 5.4 & 5.2 & 4.6 & 2.0 & 8.1 & 6.8\\
        $\gamma_p^{\prime\mathrm{break}}$, log&  -  &  -  &  - & 6.7 &  -  &  -  \\
        $\gamma_p^{\prime\mathrm{max}}$, log  & 6.1 & 7.1 & 6.9 & 8.5 & 9.2 & 8.3 \\
        $p_p^{\mathrm{low}}$                  &  -  &  -  &  -  & 0.3 &  -  &  -  \\
        $p_p^{\mathrm{high}}$                 & 1.5 & 1.5 & 1.5 & 2.3 & 2.4 & 1.5 \\
        \midrule
        $\chi^2_\mathrm{SED}$/d.o.f.                 & 0.3 & 2.7 & 1.0 & 0.7 & 3.8 & 1.6\\
        \midrule
        $N_{\mathrm{events}}$ per year              & $0.47^{+2.19}_{-0.47}$ & \phantom{.,}$3.19^{+1.90}_{-1.71}$ & $1.27^{+0.8}_{-0.55}$ & $0.02^{+0.01}_{-0.01}$ & $0.05^{+0.02}_{-0.01}$ & $0.05^{+0.02}_{-0.02}$ \\
        $N_{\mathrm{events}}$ (total)  & \phantom{,}$1.77_{-1.77}^{+8.23}$ & $10.94_{-5.84}^{+6.56}$ & $4.32_{-1.87}^{+2.71}$ & $0.07_{-0.04}^{+0.05}$ & $0.17_{-0.03}^{+0.06}$ & $0.17_{-0.04}^{+0.06}$ \\

        \bottomrule
    \end{tabular}

    \label{tab:parameters}
\end{table*}

\subsection{Expected neutrino event rates}

In this section, we estimate the number of expected neutrinos using the tabulated effective area for the point-source analysis with IceCube in its 86-string configuration and event selection applied in 2012\footnote{\url{https://icecube.wisc.edu/science/data/PS-3years}} \citep{2017ApJ...835..151A}. To emulate the conditions of the realtime stream \citep{2017APh....92...30A,Blaufuss:2019fgv} we apply an energy threshold of $>100$~TeV to the point-source effective area. At high energies the effective areas of the streams should converge\footnote{We have used the published point-source effective area of IC86 instead of using the realtime effective area or the published effective areas of the partial detector configurations, because the IC86 effective area is available with a fine declination binning of 0.01 in cosine declination, while the others distinguish only between up and down-going.}. The number of events is obtained by assuming the duration of the three different states mentioned above. We account for the fact that the detector operated with only a partial volume from August 2008 to May 2011, by scaling the IC86 effective area with the square root of the ratio of deployed strings\footnote{We expect that vertical tracks scale with the ratio of the number of strings, while horizontal tracks would scale with the square root of the number of strings. Since we mostly interested in high-energy events, the horizontal events are most relevant.} (i.e. $\sqrt{40/86}$ during the phase of operation with the 40-string configuration, IC40).

\begin{figure}
    \includegraphics[width=\linewidth]{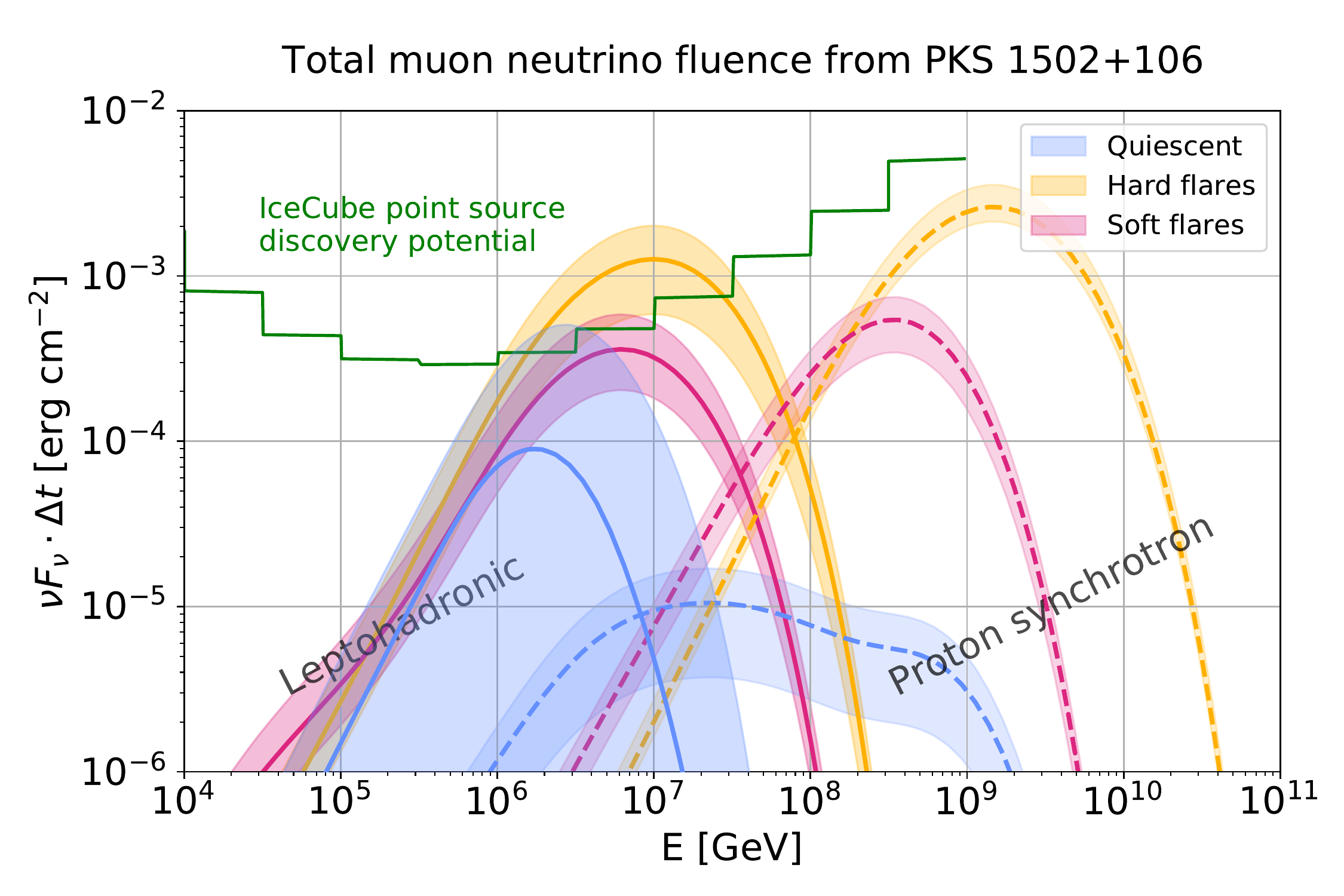}
    \caption{Total muon neutrino fluence from \PKS predicted by the leptohadronic (solid) and protons synchrotron model (dashed) in the total duration of each of the three activity states. The curves correspond to the neutrino spectra of \Fig~\ref{fig:seds}, but the flux has now been integrated over the total duration of each activity state (3.8 years quiescent, 3.7 years of hard flares, and 3.5 years of soft flares, \cf~\Tab~\ref{tab:flares}). In green we represent the IceCube discovery potential of the seven-year point source analysis for a declination of 0$^\circ$ \citep{2017ApJ...835..151A}, where the flux has been integrated over the entire seven years of that analysis. We note that the discovery potential represented here should not be used for calculating precise model predictions, since the detector was still in construction during the investigated periods and the source is at 10$^\circ$ declination. Both these aspects were considered in obtaining the numbers shown in the bottom row of \Tab~\ref{tab:parameters}.}
    \label{fig:fluence}
\end{figure}

Firstly, in order to compare the total neutrino fluence during the different activity states of the source we need to integrate the neutrino fluxes given in \Fig~\ref{fig:seds} in the total duration of each state. The result is shown in \Fig~\ref{fig:fluence}. For comparison, we show in green the differential fluence corresponding to the IceCube discovery potential at $0^\circ$ \citep{2017ApJ...835..151A}. This has been obtained by multiplying the discovery potential flux of the seven-year point source analysis with the duration of the experiment.

As we can see, during the entire course of the quiescent state, the source emits a total neutrino fluence of up to $10^{-3}\,\mathrm{erg}\,\mathrm{cm}^{-2}$ in the leptohadronic model (blue shaded area) on the left hand side. It can therefore surpass the fluence corresponding to the IceCube point-source discovery potential; however, the best fit (solid blue curve) yields only $2\times10^{-4}\,\mathrm{erg}\,\mathrm{cm}^{-2}$.

The neutrino fluence from hard flares in the leptohadronic model (solid yellow curve) surpasses the discovery potential even when its lower limit is considered. 
While the proton synchrotron model predicts a higher neutrino flux during hard flares (dashed yellow curve), the high neutrino energies place it below the discovery threshold due to absorption in the Earth, which makes the model more compatible with the lack of IceCube neutrino events during flares.
The leptohadronic model is in tension with the lack of neutrino events during the hard flaring periods. On the other hand, as mentioned above, a considerable fraction of the hard flares of \PKS took place during the construction phase of IceCube, when it was operating at partial effective volume, which lowers the overall average effective area. Furthermore, the green curve corresponds to the discover potential for a declination of 0$^\circ$, as published by the IceCube collaboration \citep{Aartsen:2014cva}, while at the declination of \PKS (10$^\circ$) the neutrino absorption in the Earth is higher, thus raising further the effective discovery potential fluence. Both these aspects are taken into account in the calculation of the predicted number of events below.

\begin{figure}
    \includegraphics[width=\linewidth]{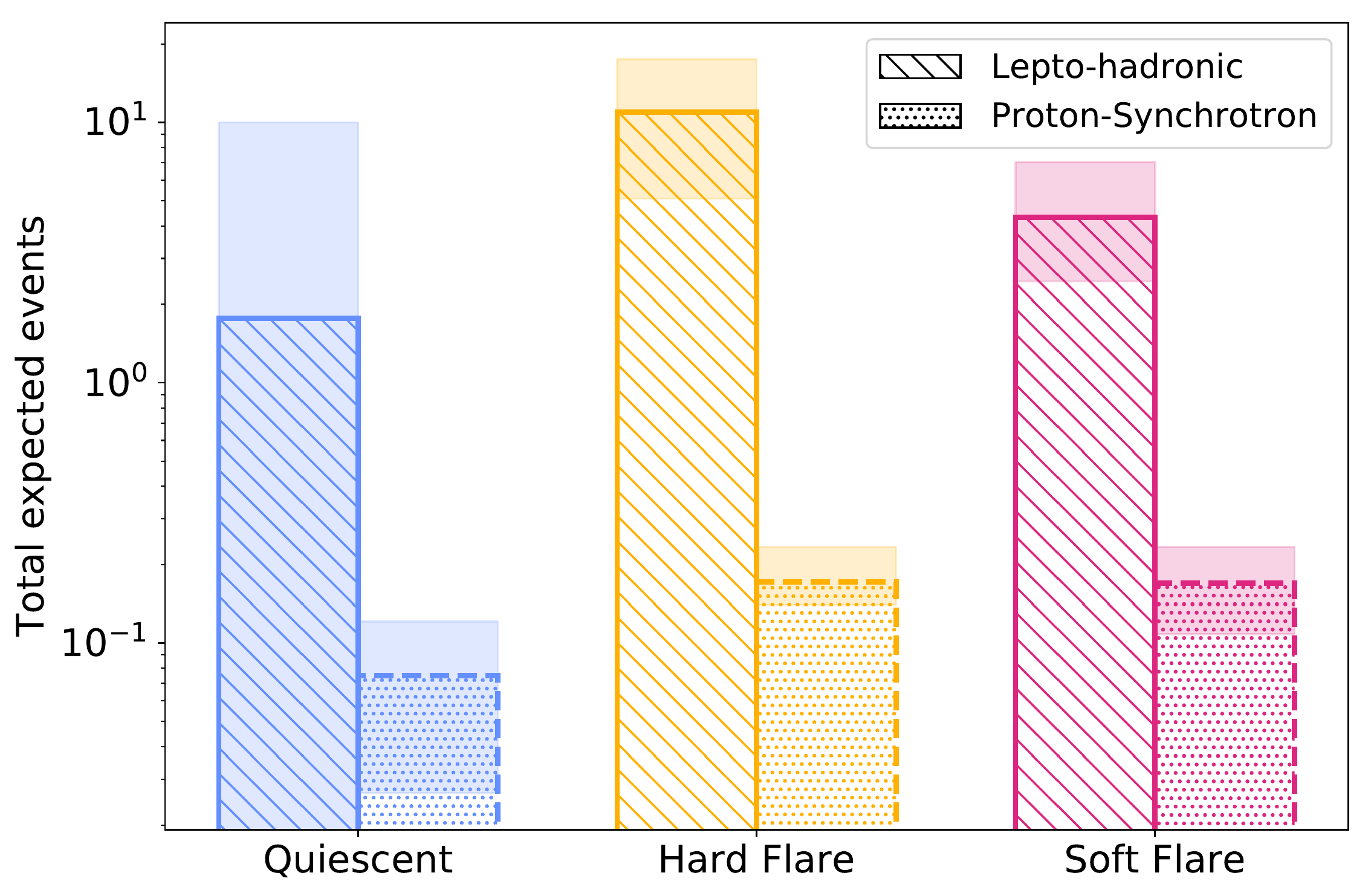}
    \caption{Total, time-integrated number of IceCube events from \PKS, expected from the leptohadronic and protons synchrotron models, in the IceCube point source analysis, in the entire duration of each the three states indicated in \Fig~\ref{fig:slots}. }
    \label{fig:n_events}
\end{figure}

In \Fig~\ref{fig:n_events} we show the total number of events in IceCube predicted during the 11 year period of \textit{Fermi}-LAT observations from \PKS, separated into the three different activity states.

As we can see, the leptohadronic model predicts a larger number of neutrino events in all cases compared to proton synchrotron. In the quiescent state, that is due to a much higher emitted neutrino flux predicted by the model. During flaring states, the leptohadronic model predicts a broader and softer spectrum that also translates into a higher number of neutrinos. 
Moreover, the effective area drops with energy above 10~PeV due to absorption in the Earth, further reducing the number of observed events predicted by the proton synchrotron model. 

The range of proton injection luminosities shown as bands in \Fig~\ref{fig:seds} translates into a systematic uncertainty in the number of IceCube events, shown as shaded regions in \Fig~\ref{fig:n_events}. In the leptohadronic model, the number of events during the quiescent state ranges from zero to a few, while in the proton synchrotron model the value lies below 0.13. The most striking difference is in the number of events during the 3.8 years of hard flares, which is 10.8 in the leptohadronic model, and only 0.17 in the proton synchrotron model. As discussed in the next section, these value ranges can be interpreted statistically, given the fact that no events were observed during flares, and one candidate neutrino from the source was observed during the quiescent state.

\section{Discussion and interpretation} \label{sec:discussion}

Both leptohadronic and proton synchrotron models can explain the high-energy emission from \PKS during the different epochs. This poses a challenge to distinguish between the two approaches. Moreover, the neutrino flux scales with the X-rays in both models, which are co-produced due to in-source cascades.

An interesting aspect of the leptohadronic model is its similarity to that applied to the 2014/15 neutrino flare of blazar \TXS by \citet{Rodrigues:2018tku} \citep[cf. also the models by][]{Reimer,Petropoulou:2019zqp}. There, the presence of external fields from a BLR leads to a high flux of MeV photons and a low-energy cutoff in the gamma-ray spectrum. In terms of the origin of the multi-wavelength fluxes, this model is also close to the interpretation by \citet{Gao:2018mnu} of the 2017 neutrino observed from \TXS: the soft X-rays (and potentially, hard gamma-rays) are interpreted as hadronic contributions, while the optical, hard X-rays and soft gamma rays are leptonic in origin. The difference is that in that case no external fields were considered. \citet{Keivani:2018rnh} have also provided a description of the multi-wavelength emission from that source during the 2017 event that did include external fields, although in that case the hadronic component was sub-dominant even in X-rays.

Regarding energetics, one aspect that is shared by both models is the particularly high minimum Lorentz factors of the accelerated protons, $\gamma_p^{\prime\mathrm{min}}$. 
The high $\gamma_{p}^{\prime\mathrm{min}}$ is necessary to comply with the X-ray constraints, in contrast to the frequently used assumptions on the particle acceleration mechanisms. Therefore, specific assumptions would be necessary for the acceleration zone, for example, assuming only protons on the high-energy end of the spectrum can leak from the acceleration zone into the radiation zone \citep{Katz:2010tv}.
This assumption can be recently found in modern interpretations of ultra-high energy cosmic rays, as it can be shown that hard cosmic-ray escape spectra from the sources are needed to describe data for sources tracing  the star formation rate, see e.g. \citet{Heinze:2019jou}; a detailed discussion of such escape mechanisms potentially producing such spectra can be found in Sec.~IIIB of \citet{Zhang:2017moz}. Similar conditions  may apply to the acceleration zone here.

Another criterion regarding energetics is the capability of the source to power the spectrum of non-thermal protons required by each model. 
To roughly estimate the plausibility of this energetic demand, the power required in protons can be compared to the Eddington luminosity of the source, which corresponds to the maximum power that can be steadily emitted by a black hole of a given mass.
The Eddington luminosity of \PKS can be estimated to be $L_\mathrm{Edd}=10^{47}(M_\mathrm{BH}/10^9M_\odot)~\mathrm{erg/s}$, where $M_\mathrm{BH}\approx10^9M_\odot$ is the black-hole mass~\citep{DElia:2002ujp,2011ApJS..194...45S}.
At the same time, both models require a proton luminosity in the rest frame of the jet of $L^{\prime}_\mathrm{p}=5\times10^{45}$-$10^{47}~\mathrm{erg/s}$. The corresponding physical luminosity (in the jet frame of the supermassive black hole) is $L^\mathrm{phys}_p = L^\prime_p \Gamma_\mathrm{b}^2/2$. Given that the Lorentz factors predicted by the model are in the range of $\Gamma_\mathrm{b}=30$-50, this yields a physical luminosity in protons in the range of 20-200 $L_\mathrm{Edd}$ for all the models.
This figure is not significantly reduced during the quiescent state compared to the flares, if the best-fit parameters in \Tab~\ref{tab:parameters} are considered. Although the Eddington luminosity does not set a hard limit on the available proton power, we can probably disfavor scenarios where the source must accelerate protons in a super-Eddington regime during long quiescent states. This energetic `crisis' of proton synchrotron blazars has already been noted by \citet{Liodakis:2020dvd}, who studied a large sample of blazars including \PKS.

The latter observation certainly argues in favor of the leptohadronic model, where the quiescent state is compatible with an electron-only or electron-dominated emission. The quiescent states of the source would then be mostly dominated by leptonic emission (for example between 2010 and 2015, \cf~\Fig~\ref{fig:slots}). Sporadically, a neutrino-efficient state can be achieved through a temporary increase in the proton injection power, with all other parameters unaltered, without significant changes in the UV, X-ray, or gamma-ray fluxes (blue shaded area in \Fig~\ref{fig:seds}, left-hand plot).

Regarding the number of IceCube events, we concluded that the proton synchrotron model predicts only between 0.03 and 0.17 neutrinos for each activity state. This would be consistent with seeing one neutrino from the source during the flaring state, if we assume that this is not the only source of this type in the Universe \citep[i.e., by accounting for the Eddington bias, as discussed by][]{Strotjohann:2018ufz}.
However, the argument can also be made that the emission of blazars can often be well described by a leptonic model, while in this work we describe the flares of \PKS with a non-negligible hadronic contribution. This would place the source in a `special' sub-class of neutrino-efficient blazars that may be considerably more reduced compared to the total blazar population. Such an argument would then lead to a weaker effect of the Eddington bias.

In the case of the leptohadronic model the neutrino predictions are considerably higher, $1.77_{-1.77}^{+8.23}$ events predicted during the quiescent state (\cf~\Fig~\ref{fig:n_events} and \Tab~\ref{tab:parameters}). This number is in agreement with the observation of one event by IceCube. On the other hand, the model also predicts a high number of events during flaring states, especially hard flares. In total $10.94^{-5.84}_{+6.56}$ events are expected, of which $7.97^{-4.25}_{+4.78}$ in the IC86 period, while either the IceCube realtime system was operational \citep[starting from 2016,][]{2017APh....92...30A} or an archival search was applied \citep{TXS_MM}. This means that our model predicts a minimum of 7.97-4.25=3.72 high-energy events detected in IceCube during hard flares, which is in mild tension with the non-detection of any events (p-value of 2.4\%, assuming Poisson statistics). We therefore strongly encourage a similar search for high-energy events during the flaring period in 2009: if also during that period no high-energy neutrino is found, the leptohadronic model would be disfavored much more significantly, with a p-value of $6 \cdot 10^{-3}$.

Finally, we note that \citet{Aartsen:2019fau} have set a stringent 90\% C.L. neutrino flux upper limit on \PKS of $2.6 \times 10^{-13} (E_\nu / \mathrm{TeV})^{-2}\,\mathrm{TeV}^{-1}\,\mathrm{cm}^{-2}\,\mathrm{s}^{-1}$. However, this limit was set assuming a power-law spectrum over a broad energy range, which is very different from the predicted spectral shapes by our models and can therefore not be directly compared to our results.

\section{Summary and conclusions} \label{sec:conclusion}

We have interpreted the multi-epoch, multi-wavelength observations of \PKS, one of the brightest gamma-ray blazars detected. Using one-zone radiation models we have estimated the possible range of neutrino spectra emitted by the source while self-consistently explaining the multi-wavelength emission during its different states of electromagnetic activity: the quiescent state, and flares with a hard and with a soft gamma-ray spectrum. We have focused on the emission in the range from ultraviolet to gamma rays, in their different flux levels over the 11 years of \textit{Fermi}-LAT observations. We have found that the emission can be well described with a hadronic contribution, both during quiescent and flaring states.  In addition to best-fit results, we also provided an uncertainty range in the luminosity of non-thermal protons, which results in a systematic uncertainty in the neutrino emission level. 

The X-ray fluxes observed during flares, which are typically bright and with a soft spectrum, are difficult to explain by means of a purely leptonic one-zone model. In both these models, this X-ray emission is a sub-dominant contribution originating in electromagnetic cascades initiated by hadronic interactions in the source. This point seems to support proton acceleration in \PKS independently of neutrino emission.

Regarding gamma-ray emission, in the lephohadronic model, it is most often dominated by inverse Compton scattering of photons from a broad line region. In the proton synchroton model, gamma rays below 1 GeV originate in proton synchrotron emission, and those above 1 GeV originate from electron/positron pairs produced by those same protons. On the other hand, the archival radio observations from the source cannot originate in a region as compact as the one modeled in this work, because of its radio-opacity due to synchrotron self-absorption. This suggests a larger dissipation region in the jet with efficient electron synchrotron emission. It is noteworthy that the bright radio emission from the source in the years leading up to the neutrino emission seems to point towards a correlation (or possibly even a common origin) between radio and neutrino emission. This point is driven further by statistical correlations found between radio-loud AGN and the observed high-energy neutrinos \citep{Plavin:2020emb,Hovatta:2020lor}. However, our extensive scan of the source's parameter space has revealed that such common origin is not feasible within a one-zone model, and other more complex geometries may be necessary to potentially establish such a bridge.

We have then drawn conclusions on the viability of the two models based on energetics and their predictions for the neutrino flux. The main difference between the two models is that the proton synchrotron model requires constant acceleration of protons, even during the quiescent state, in order to explain X-ray and gamma-ray observations. This implies a constant neutrino output from the source, which is statistically below the IceCube sensitivity. At the same time, the model demands a super-Eddington power in non-thermal protons during the entire quiescent state, which presents a major challenge from the energetic point of view.

On the other hand, a leptohadronic model is compatible with a purely leptonic solution during quiescent states. This suggests a quiescent state that is dominated by electron emission most of the time, but is also compatible with periods of efficient neutrino emission through temporary increases in the injection of non-thermal protons without significant changes in the multi-wavelength emission. This can help explain the IceCube event 190730A from the direction compatible with the position of the source. Moreover, similar models have been used in the literature to explain the multi-wavelength emission from blazar \TXS during the 2017 neutrino event. The limitation of the leptohadronic model resides in the high number of IceCube events ($7.97^{-4.25}_{+4.78}$) predicted during hard gamma-ray flares since the beginning of the realtime alert system. This result is in mild tension with the non-observation of neutrino events during flares. An archival search of IceCube events from the direction of the source before 2010 could help constrain more significantly this leptohadronic model: if no events were found, the model would be excluded with a p-value of $6\cdot 10^{-3}$.

{\bf Acknowledgments.}  We would like to thank Robert Stein and Anatoli Fedynitch for fruitful discussions and to Julia Tjus for comments on the manuscript. This work has been supported by the Initiative and Networking Fund of the Helmholtz Association and by the European Research Council (ERC) under the European Unions Horizon 2020 research and innovation programme (Grant No. 646623).


\end{document}